\documentclass[conference, anonymous=true]{article}

\usepackage{biblatex}
\usepackage[nonatbib, final]{neurips_2022}

\usepackage[utf8]{inputenc} 
\usepackage[T1]{fontenc}    
\usepackage{hyperref}       
\usepackage{url}            
\usepackage{booktabs}       
\usepackage{amsfonts}       
\usepackage{nicefrac}       
\usepackage{microtype}      
\usepackage{xcolor}         

\title{On the notion of Hallucinations from the lens of Bias and Validity in Synthetic CXR Images}

\author{%
Gauri Bhardwaj$^{1,3}$ \quad Yuvaraj Govindarajulu$^{1}$ \quad Sundaraparipurnan Narayanan$^2$ \\
\quad \textbf{Pavan Kulkarni}$^1$ \quad \textbf{Manojkumar Parmar}$^1$ \\
$^1$AIShield, Bosch Global Software Technologies \quad $^2$AI Tech Ethics \quad $^3$Manipal Institute of Technology\\
\texttt{gaurisanjeevbhardwaj@gmail.com}\\
\texttt{sundar.narayanan@aitechethics.com}\\
\texttt{\{govindarajulu.yuvaraj,pavan.kulkarni,manojkumar.parmar\}@in.bosch.com}
}

\begin{document}

\maketitle

\begin{abstract}
 Medical imaging has revolutionized disease diagnosis, yet the potential is hampered by limited access to diverse and privacy-conscious datasets. Open-source medical datasets, while valuable, suffer from data quality and clinical information disparities. Generative models, such as diffusion models, aim to mitigate these challenges. At Stanford, researchers explored the utility of a fine-tuned Stable Diffusion model (RoentGen) for medical imaging data augmentation.
Our work examines specific considerations to expand the Stanford research question, “Could Stable Diffusion Solve a Gap in Medical Imaging Data?” from the lens of bias and validity of the generated outcomes. We leveraged RoentGen to produce synthetic Chest-XRay (CXR) images and conducted assessments on bias, validity, and hallucinations. Diagnostic accuracy was evaluated by a disease classifier, while a COVID classifier uncovered latent hallucinations. The bias analysis unveiled disparities in classification performance among various subgroups, with a pronounced impact on the Female Hispanic subgroup. Furthermore, incorporating race and gender into input prompts exacerbated fairness issues in the generated images. The quality of synthetic images exhibited variability, particularly in certain disease classes, where there was more significant uncertainty compared to the original images. Additionally, we observed latent hallucinations, with approximately 42\% of the images incorrectly indicating COVID, hinting at the presence of hallucinatory elements. These identifications provide new research directions towards interpretability of synthetic CXR images, for further understanding of associated risks and patient safety in medical applications. 

\end{abstract}

\section{Introduction}

Medical imaging technologies have enabled precise disease diagnosis without invasive tests, yet radiologists lack access to large amounts of training images due to privacy and security concerns \cite{Loten2020eraofai} \cite{Celard2022DLImg}. The lack of high-quality annotated medical imagery datasets provides the impetus for generative imaging models to be created that accurately represent medical concepts while providing compositional diversity \cite{chambon2022roentgen}. The utilization of open-source medical image datasets presents numerous limitations, which include inconsistencies in image quality, gaps in metadata and clinical information, the use of outdated imaging equipment in some dataset acquisitions, the presence of low-quality images, insufficient expert labeling, and often, restrictions on commercial usage imposed by many open-source datasets\cite{willem2020prepareimage}. Some performance variations could be due to differences in patient demographics across sites, and the study highlights the need for further investigation in this area. For instance, a study found that the AI algorithm’s performance dropped substantially when evaluated on a different site, indicating limitations in generalizability \cite{wu2021medical}. 
Deep learning models can create synthetic data to improve diagnostic accuracy and advance medical imaging techniques\cite{nie2018medImg}. The diffusion model has gained popularity as it better estimates data distribution and sample quality than other methods - making it a potential alternative to GANs when creating images \cite{nguyen2023imggen}. Today, advanced models like Stable Diffusion\cite{rombach2022highresolution}, DALL-E 2, and Midjourney are now capable of producing impressive-quality images from natural language descriptions.\cite{Aliborji2023generatedFaces}\cite{rando2022redteaming}.

Multi-modal models trained on natural image/text pairs often perform poorly when applied to medical applications \cite{chambon2022roentgen}. Typically, diffusion models follow a two-stage process: (1) a prior generating CLIP image embedding with a text caption; (2) a decoder (U-net) producing an image conditioned upon the image embedding. In this context, diffusion models offer advantages as decoders due to their computational efficiency and ability to generate high-quality samples \cite{ramesh2022hierarchical-CLIP}.

\section{Stable Diffusion applied to medical imaging}

While the license agreements from Stable Diffusion limit its use for medical advice or interpretation, researchers at Stanford used the technique to generate radiology images (RoentGen)\cite{chambon2022roentgen} to assess if this method can solve the current gap of lack of adequate data in medical imaging. With RoentGen, the researchers could generate realistic synthetic CXR images using free-form text prompts, improving diagnostic accuracy. They found that fine-tuning the model on a fixed training set increases classifier performance by 5\% when trained on both synthetic and original images and 3\% when trained solely on synthetic images. Furthermore, text-encoder representation capabilities for certain diseases like Pneumothorax were enhanced by 25\% through fine-tuning.

The RoentGen study investigated using latent diffusion models to generate medical images in thoracic imaging. The researchers list the limitations encountered, such as difficulty measuring clinical accuracy with standard metrics (e.g., usefulness of image), lack of diversity in generated images (only a small sample considered), and need for improved text prompts (e.g., radiology-specific words or verbatim radiology text). The study or the paper did not discuss the known limitations of medical imaging interpretation, including data diversity, bias, validity, and interpretability of the underlying model performance.  
Our work examines specific considerations to expand the Stanford research question, “Could Stable Diffusion Solve a Gap in Medical Imaging Data?” \cite{nikki2022stablediff} from the lens of bias and validity of the generated outcomes.

\section{Experimental Methodology for validating Bias, Adversarial Robustness and Reliability
}
We rigorously validated our dataset and assessed RoentGen's performance, initializing it with model weights from its creators. Our primary data source was the MIMIC-CXR dataset, comprising CXR images and radiologist textual inputs \cite{mimic-Net}. An experimental framework was then established, utilizing pre-trained RoentGen model weights, ensuring a robust foundation for image generation. We subsequently generated CXR images based on radiologist textual descriptions from the MIMIC-CXR dataset, with and without protected variables such as Gender and Race as counterfactuals.
To gauge image quality and fidelity to medical conditions, we employed a disease classification model trained on multiple- CXR illness categories \cite{cohen2020chester}. The classifier's performance on original and generated images was compared, providing a comprehensive assessment of the model's representational capabilities. Additionally, we probed for latent hallucinations by subjecting both original and generated images to a dedicated COVID classifier \cite{LindaWang2020COVID-Net}. As the original images were not expected to show signs of COVID, any discrepancies in the classifier's results signaled potential features introduced by RoentGen's hallucinations. We then rigorously compiled and analyzed the results from our experiments.

\section{Results}

\subsection{Bias and Fairness}
The RoentGen model was employed to construct a synthetic CXR dataset, which was subsequently compared to its MIMIC counterpart (labelled ground truth) to assess their performance with regard to fairness and bias metrics. After filtering out scans with consistent demographic information and retaining one scan per patient, we arrived at a subset of MIMIC images comprising 18,200 samples. Inference was conducted on both the MIMIC images and the synthetically generated images using the all-weights model from the TorchXRayVision library \cite{cohen2021torchxrayvision}. To examine fairness and bias, subgroups of gender and ethnicity were analyzed and graded based on indicators stressing True Positive Rate (TPR) and Selection Rate (SR) disparity. The findings indicate that for the disease Atelectasis, the model displayed bias against specific subgroups, with a notable and enduring bias observed in the Hispanic female subgroup (which was also concluded in the CheXclusion study \cite{seyyedkalantari2020chexclusion}). Upon examining TPR and SR values, the model consistently exhibits bias against the Hispanic female subgroup, which also demonstrates poorer performance compared to other subgroups despite its larger sample size. This bias persists in synthetic images, with TPR values dropping to 44\% for Hispanic females, in contrast to 64\% for White females and 56\% for Asian females. The introduction of racial and gender references in prompts further exacerbates bias, resulting in a substantial 26\% deviation from the SR of White females and a 16\% deviation from Asian females. Consequently, the synthetic images under-perform in comparison to their MIMIC counterparts, with demographic references intensifying the bias.

\subsection{Validity}
 We also observed differences in the characteristics of the original image and the synthetic image generated from RoentGen, indicating possible generalization of features in the synthetic images during generation. To understand the extent of generalization and potential implications therein, we evaluated the quality of the generated images by passing them through a classification model (Chester, the AI Radiology Assistant \cite{cohen2020chester}) trained on a comprehensive set of (18-classes) illness categories. This was done to accurately assess the model's ability to capture and represent the medical conditions. About half of the above mentioned categories had an accuracy of 50-70\%, while the rest had an accuracy of more than 70\% when compared to the performance of original images using the same classifier. The results showed lower confidence (high uncertainty) towards selected classes for the synthetic images despite the classifier's high accuracy. The synthetic samples were also more prone to false negative classifications compared to their MIMIC counterparts. Higher uncertainty indicates that the samples are closer to the decision boundaries and have a potential for being misclassified with smaller perturbations.
\subsection{Hallucinations}

To further understand if the differences in quality arising from the generalization of features in synthetic image generation represent latent hallucinations, we passed both the generated and original images through a dedicated COVID classifier \cite{LindaWang2020COVID-Net}. The results showed that RoentGen adds additional latent features (hallucinations) in the generated synthetic images that do not represent attributes or correspond to the original text descriptions provided as prompts. In this case, the latent features represented COVID in 7500 out of 18,200 images, highlighting hallucinations. This also exhibits the need to evaluate the generated images against classes independent of the training set to understand the hallucinations. In our experiment, when passing the generated images through the classifier containing the original class, the accuracy was better for all the classes (while there were uncertainties in confidence); however, when passing the generated images through the classifier containing the (out-of-training) independent class, the images exposed, potential latent hallucinations also resulting in lower classification accuracy.

\section{Discussion and Outlook}

Interpretation of Generative AI models is challenging and newer methods such as per-word attribution \cite{tang2022daam} or per-head cross attention \cite{hertz2022prompttoprompt} have been proposed. It is essential to have an extensive system for comprehending text-to-image attribution to detect any potential flaws or inconsistencies within these models in medical imaging. The harms contributed by the inherent biases in these pre-trained models may have accentuated implications considering the general lack of adequate interpretability of these models. Given the risks noticed from the experimental results, we believe there is a need to have more comprehensive approaches for evaluating the suitability of using Generative models for data augmentation in disease diagnosis.

The specific aberrations arising out of hallucinations from our experimental study were not identified through traditional evaluation metrics for generative images, thereby requiring domain-specific consideration of alternative methods. We choose an out-of-training class validation method to demonstrate the hallucination; however, such an approach may not be suited for many circumstances. We believe these experimental results will trigger further research on establishing validity and reliability metrics for hallucinations in Generative Medical Images. It is also pertinent to consider the need for extended research on approaches to limit latent hallucinations, as some of these hallucinations, if not constrained, can have significant downside impacts on people and the planet.





\printbibliography

@misc{ramesh2022hierarchical-CLIP,
      title={Hierarchical Text-Conditional Image Generation with CLIP Latents}, 
      author={Aditya Ramesh and Prafulla Dhariwal and Alex Nichol and Casey Chu and Mark Chen},
      year={2022},
      eprint={2204.06125},
      archivePrefix={arXiv},
      primaryClass={cs.CV}
}

@misc{Aliborji2023generatedFaces,
      title={Generated Faces in the Wild: Quantitative Comparison of Stable Diffusion, Midjourney and DALL-E 2}, 
      author={Ali Borji},
      year={2023},
      eprint={2210.00586},
      archivePrefix={arXiv},
      primaryClass={cs.CV}
}

@article{wu2021medical,
  title={How medical AI devices are evaluated: limitations and recommendations from an analysis of FDA approvals},
  author={Wu, Eric and Wu, Kevin and Daneshjou, Roxana and Ouyang, David and Ho, Daniel E and Zou, James},
  journal={Nature Medicine},
  volume={27},
  number={4},
  pages={582--584},
  year={2021},
  publisher={Nature Publishing Group US New York}
}

@article{nie2018medImg,
  title={Medical image synthesis with deep convolutional adversarial networks},
  author={Nie, Dong and Trullo, Roger and Lian, Jun and Wang, Li and Petitjean, Caroline and Ruan, Su and Wang, Qian and Shen, Dinggang},
  journal={IEEE Transactions on Biomedical Engineering},
  volume={65},
  number={12},
  pages={2720--2730},
  year={2018},
  publisher={IEEE}
}

@misc{LindaWang2020COVID-Net,
      title={COVID-Net: A Tailored Deep Convolutional Neural Network Design for Detection of COVID-19 Cases from Chest X-Ray Images}, 
      author={Linda Wang, Zhong Qiu Lin and Alexander Wong},
      year={2020},
      eprint={2003.09871},
      archivePrefix={arXiv},
      primaryClass={eess.IV},
      URL={https://arxiv.org/pdf/2003.09871.pdf}
}

@article{Celard2022DLImg,
author = {Celard, Pedro and Iglesias, Eva and Sorribes-Fdez, J. and Romero, Rubén and Vieira, Seara and Borrajo, María},
year = {2022},
month = {11},
pages = {1-33},
title = {A survey on deep learning applied to medical images: from simple artificial neural networks to generative models},
volume = {35},
journal = {Neural Computing and Applications},
doi = {10.1007/s00521-022-07953-4}
}

@misc{chambon2022roentgen,
      title={RoentGen: Vision-Language Foundation Model for Chest X-ray Generation}, 
      author={Pierre Chambon and Christian Bluethgen and Jean-Benoit Delbrouck and Rogier Van der Sluijs and Małgorzata Połacin and Juan Manuel Zambrano Chaves and Tanishq Mathew Abraham and Shivanshu Purohit and Curtis P. Langlotz and Akshay Chaudhari},
      year={2022},
      eprint={2211.12737},
      archivePrefix={arXiv},
      primaryClass={cs.CV}
}

@misc{nikki2022stablediff,
      title={Could Stable Diffusion Solve a Gap in Medical Imaging Data}, 
      author={Nikki Goth Itoi},
      url={https://hai.stanford.edu/news/could-stable-diffusion-solve-gap-medical-imaging-data},
      year={2022}
}

@misc{mimic-Net,
      title={MIMIC-CXR, a de-identified publicly available database of chest radiographs with free-text reports}, 
      author= {Alistair E. W. Johnson et. al},
      year={2019},
      DOI = {10.3390/app10175729},
      URL = {https://physionet.org/content/mimic-cxr/2.0.0/}
}

@misc{hertz2022prompttoprompt,
      title={Prompt-to-Prompt Image Editing with Cross Attention Control}, 
      author={Amir Hertz and Ron Mokady and Jay Tenenbaum and Kfir Aberman and Yael Pritch and Daniel Cohen-Or},
      year={2022},
      eprint={2208.01626},
      archivePrefix={arXiv},
      primaryClass={cs.CV}
}

@misc{Loten2020eraofai,
      title={Medical Imaging and Privacy in the Era of Artificial Intelligence: Myth, Fallacy, and the Future}, 
      author={Eyal Lotan and Charlotte Tschider and Daniel K Sodickson and Arthur L Caplan and Mary Bruno and Ben Zhang and Yvonne W Lui},
      year={2020},
      doi={0.1016/j.jacr.2020.04.007}
}

@misc{cohen2021torchxrayvision,
      title={TorchXRayVision: A library of chest X-ray datasets and models}, 
      author={Joseph Paul Cohen and Joseph D. Viviano and Paul Bertin and Paul Morrison and Parsa Torabian and Matteo Guarrera and Matthew P Lungren and Akshay Chaudhari and Rupert Brooks and Mohammad Hashir and Hadrien Bertrand},
      year={2021},
      eprint={2111.00595},
      archivePrefix={arXiv},
      primaryClass={eess.IV}
}

@misc{cohen2020chester,
      title={Chester: A Web Delivered Locally Computed Chest X-Ray Disease Prediction System}, 
      author={Joseph Paul Cohen and Paul Bertin and Vincent Frappier},
      year={2020},
      eprint={1901.11210},
      archivePrefix={arXiv},
      primaryClass={cs.CV}
}

@misc{seyyedkalantari2020chexclusion,
      title={CheXclusion: Fairness gaps in deep chest X-ray classifiers}, 
      author={Laleh Seyyed-Kalantari and Guanxiong Liu and Matthew McDermott and Irene Y. Chen and Marzyeh Ghassemi},
      year={2020},
      eprint={2003.00827},
      archivePrefix={arXiv},
      primaryClass={cs.CV}
}

@inproceedings{nguyen2023imggen,
title = "A New Chapter for Medical Image Generation: The Stable Diffusion Method",
author = "Nguyen, {Loc X.} and Aung, {Pyae Sone} and Le, {Huy Q.} and Park, {Seong Bae} and Hong, {Choong Seon}",
year = "2023",
doi = "10.1109/ICOIN56518.2023.10049010",
language = "English",
series = "International Conference on Information Networking",
publisher = "IEEE Computer Society",
pages = "483--486",
booktitle = "37th International Conference on Information Networking, ICOIN 2023",
address = "United States",
}

@misc{rando2022redteaming,
      title={Red-Teaming the Stable Diffusion Safety Filter}, 
      author={Javier Rando and Daniel Paleka and David Lindner and Lennart Heim and Florian Tramèr},
      year={2022},
      eprint={2210.04610},
      archivePrefix={arXiv},
      primaryClass={cs.AI}
}

@misc{rombach2022highresolution,
      title={High-Resolution Image Synthesis with Latent Diffusion Models}, 
      author={Robin Rombach and Andreas Blattmann and Dominik Lorenz and Patrick Esser and Björn Ommer},
      year={2022},
      eprint={2112.10752},
      archivePrefix={arXiv},
      primaryClass={cs.CV}
}

@misc{tang2022daam,
      title={What the DAAM: Interpreting Stable Diffusion Using Cross Attention}, 
      author={Raphael Tang and Linqing Liu and Akshat Pandey and Zhiying Jiang and Gefei Yang and Karun Kumar and Pontus Stenetorp and Jimmy Lin and Ferhan Ture},
      year={2022},
      eprint={2210.04885},
      archivePrefix={arXiv},
      primaryClass={cs.CV}
}

@misc{willem2020prepareimage,
      title={Preparing Medical Imaging Data for Machine Learning.}, 
      author={Martin J Willemink and Wojciech A Koszek and Cailin Hardell and Jie Wu and Dominik Fleischmann and Hugh Harvey and Les R Folio and Ronald M Summers and Daniel L Rubin and Matthew P Lungren},
      year={2020},
      doi={10.1148/radiol.2020192224}
}

\section*{Potential negative impacts}

Our research presents a comprehensive exploration of the drawbacks associated with the utilization of generative AI models in the context of medical imaging. Specifically, our study sheds light on issues related to bias, validity, and the phenomenon of hallucination within the datasets generated by these AI models.

It is important to acknowledge that our study, while serving as a reminder of the longstanding problem of fairness in the realm of medical AI, may inadvertently contribute to the perpetuation of skepticism surrounding the applicability of diffusion models in the medical domain. This skepticism may erode the credibility of diffusion models as a viable and convenient method for enhancing medical image datasets, potentially dissuading other researchers in the field from exploring these avenues.

Nevertheless, in light of the ethical imperatives inherent to the use of AI in medical applications, our research also serves as an illustrative example of the notion that reliance solely on generated datasets to bridge existing gaps may not yet be a sufficiently robust approach.


\end{document}